\documentclass[conference]{IEEEtran}
\IEEEoverridecommandlockouts

\usepackage[T1]{fontenc}
\usepackage{graphicx}
\usepackage{subfig}
\usepackage{multirow}
\usepackage{bm}
\usepackage{adjustbox }
\usepackage{scrextend}
\usepackage[normalem]{ulem}
\usepackage{hyperref}

\usepackage{amsmath,amsfonts,bm}

\def\eqref#1{equation~\ref{#1}}

\def\1{\bm{1}}

\def\vp{{\bm{p}}}

\DeclareMathAlphabet{\mathsfit}{\encodingdefault}{\sfdefault}{m}{sl}
\SetMathAlphabet{\mathsfit}{bold}{\encodingdefault}{\sfdefault}{bx}{n}

\def\gI{{\mathcal{I}}}

\def\gU{{\mathcal{U}}}

\def\sP{{\mathbb{P}}}

\newcommand*{\norm}[1]{\left\|#1\right\|}

\usepackage{todonotes}

\usepackage{color, soul}
\usepackage[printonlyused]{acronym}
\acrodef{AR}{auto-regressive}
\acrodef{AoA}{angle of arrival}
\acrodef{ARMA}{auto-regressive moving average}
\acrodef{CP}{cyclic prefix}
\acrodef{CDL}{clustered delay line}
\acrodef{CSI}{channel state information}
\acrodef{CIR}{channel impulse response}

\acrodef{KF}{Kalman Filter}
\acrodef{MMSE}{Minimum Mean Square Error}
\acrodef{MNSE}{Mean Normalized Square Error}
\acrodef{NSE}{Normalized Square Error}
\acrodef{NN}{Neural Network}
\acrodef{HKF}{Hypernetwork Kalman Filter}
\acrodef{BKF}{Binned Kalman Filter}
\acrodef{GKF}{genie \ac{KF}}
\acrodef{UE}{User Equipment}
\acrodef{ISI}{Intersymbol Interference}
\acrodef{LOS}{line of sight}

\acrodef{ToF}{time of flight}
\acrodef{TDoA}{time difference of arrival}

\begin{document}
\title{
Neural RF SLAM for unsupervised positioning and mapping with channel state information
}
\author{
    \IEEEauthorblockN{Shreya Kadambi\IEEEauthorrefmark{2}, Arash Behboodi\IEEEauthorrefmark{1}, Joseph B. Soriaga\IEEEauthorrefmark{2}, Max Welling\IEEEauthorrefmark{3},
    }
        \IEEEauthorblockN{
    Roohollah Amiri\IEEEauthorrefmark{2}, Srinivas Yerramalli\IEEEauthorrefmark{2}, Taesang Yoo\IEEEauthorrefmark{2}
    }
    \IEEEauthorblockA{\IEEEauthorrefmark{1}Qualcomm Technologies Netherlands B.V. 
    \IEEEauthorrefmark{2}Qualcomm Technologies, Inc.
    \IEEEauthorrefmark{3}University of Amsterdam
    }
    \IEEEauthorblockA{Qualcomm AI Research}
\thanks{Qualcomm AI Research is an initiative of Qualcomm Technologies, Inc.}
}

\maketitle

\begin{abstract}
We present a neural network architecture for jointly learning user locations and environment mapping up to isometry, in an unsupervised way, from \ac{CSI} values with no location information. The model is based on an encoder-decoder architecture. The encoder network maps \ac{CSI} values to the user location. The decoder network models the physics of propagation by parametrizing the environment using virtual anchors. It aims at reconstructing, from the encoder output and virtual anchor location, the set of \acp{ToF} that are extracted from \ac{CSI} using super-resolution methods. The neural network task is set prediction and is accordingly trained end-to-end. The proposed model learns an interpretable latent, i.e., user location, by just enforcing a physics-based decoder. It is shown that the proposed model achieves sub-meter accuracy on synthetic ray tracing based datasets with single anchor SISO setup while recovering the environment map {up to 4cm median error in a 2D environment and 15cm in a 3D environment}.

\end{abstract}

\section{Introduction}

Precise indoor positioning at sub-meter level accuracy is one of the main requirements in future communication networks such as 5G Advanced and 6G. The problem is classical with many solutions already  available. Ranging methods based on \acf{ToF}, \ac{TDoA} or \ac{AoA} require multiple anchors to find user location based on triangulation or trilateration. On the other hand, fingerprinting methods find the user location from traces like \ac{CSI} by training a supervised learning algorithm using labeled data. Ranging methods are line-of-sight dependent. Besides, \ac{ToF}  and \ac{TDoA} based methods require additional overhead for synchronization and calibration. Fingerprinting methods need labelled data, which should be consistently updated with environment change.

However, multi-path environments provide additional sources that can be leveraged for positioning. The idea is that localization is possible even with as few as one anchor, if the environment provides sufficient multi-path components at each point. In other words, the multi-path signature of each location is unique and can be used for positioning. \ac{CSI}-based fingerprinting methods are implicitly built on top of this assumption. To explicitly utilize multi-path for positioning, we need the environment map, which is also important for joint sensing and communication applications. The mapping step involves some challenges. First, we need to find a suitable way for representing the environment map, that is easy to learn from data and to use for positioning. Next, the environment map should be obtained in an efficient way without incurring additional field survey cost.

In this paper, we propose a neural network architecture that jointly learns the environment map and user locations from \ac{CSI} values without any location information about users or anchor. \ac{CSI} values are used for communication purpose and therefore available in the field. They are continuously gathered and therefore can be used to closely track environment changes. We parametrize the environment using virtual anchors and learn their location jointly with a positioning neural network. The architecture incorporates physics of propagation using \acp{ToF} extracted from \ac{CSI} values. It is shown that the proposed model can learn the position and map up to isometry with sub-meter accuracy. The paper is organized as follows. In Section \ref{sec:prob_setup}, we present the problem formulation. Our proposed neural network architecture is presented in Section \ref{sec:neural_rf_slam}. Experiment results are shown in Section \ref{sec:experiments}.

\subsection{Related Works}

The simultaneous localization and mapping (SLAM) problem asks if it is possible for a mobile robot to be placed at an unknown location in an unknown environment and for the robot to incrementally build a consistent map of this environment while simultaneously determining its location within this map~\cite{art_white_0}. SLAM problem which originally started from robotics has found its way in many fields including wireless communications with applications such as precise positioning. In particular, SLAM procedures have been successfully translated into multi-path assisted positioning methods with anchors as equivalence of landmarks/features. Many of probabilistic solutions to SLAM such as EKF-SLAM, Fast-SLAM, or Graph-SLAM have been used in multipath assisted positioning such as works in~\cite{gentner_multipath_2016, art_leitinger_BP, art_hough}. Furthermore, SLAM based positioning in wireless communications differs from RF-fingerprinting in explicitly modeling the multipath environment while RF-fingerprinting creates a correlation model between channel measurements and user locations.


Extracting accurate multipath components from channel measurements (CSI) is essential and distinguishing factor in  SLAM solutions to multipath assisted positioning. In that regard, existing works can be categorized based on their operating frequency and bandwidth such as ultra-wide band (UWB)~\cite{art_leitinger_BP, art_A035}, Sub-6GHz~\cite{gentner_multipath_2016}, and millimeter wave (mmWave) systems~\cite{art_A024, art_A047}. Estimation accuracy of multipath components is directly related to the available bandwidth or the resolution of the wireless channel. Therefore, the estimated multipath delays have higher accuracy is UWB systems while the same factor becomes a limitation in Sub-6GHz systems. Another categorization of SLAM can be based on the type of multipath components. Multipath components that only experience reflections on their propagation path are assured to carry geometrical information of the environment. It is more challenging to employ diffracted and scattered paths for SLAM, and therefore, robustness to undesired multipath components is essential.

It is more challenging to extract virtual anchor locations when only one the modalities among \ac{ToF}, range measurements, or \ac{AoA} of multipath components is available. In these cases,  multiple modes naturally arise in the distributions of locations after multiple measurements, many methods include means for maintaining multiple hypotheses~\cite{art_A079, art_A080, art_A081}. Authors in~\cite{art_A079} propose to use polar coordinates to model range-only SLAM with advantages of  better handling of multimodal distribution, no requirement on batch process for linearization and no prior needed as in numbers of virtual anchors. In~\cite{art_A080}, SLAM is formulated as a matrix factorization problem, where features of observations are linearly related to multi-dimensional state space. Spectral system identification is used to learn dynamical system parameters such as a state space, motion model, and observation model directly. Authors in~\cite{art_A081} propose relative over-parametrized EKF and polar coordinates. Machine learning for positioning has been mostly focused on supervised setting and fingerprinting  approaches. These methods range from pure neural network solutions to careful feature engineering with kNN (see \cite{Ayyalasomayajula2020DeepLB} and references therein).
\ac{CSI} is a common input to all these models. 

In contrast to fingerprinting methods, our approach is unsupervised. Our approach is a lightweight unsupervised model that utilizes multipaths undergoing multiple reflections. The approach can be potentially extended to other effects like diffraction. Our approach does not require prior data association labels between virtual anchors and extracted \acp{ToF}. The underlying network can extract spatial information given any input such as \ac{CSI} or TDoA or ToF if available. Although the model is presented for a single anchor scenario, it can be extended to multi-anchor by concatenating multiple decoders, one per anchor.

\section{Background and Problem Setup}
\label{sec:prob_setup}
\subsection{Multi-path-aided positioning}
We consider a propagation environment with a static anchor. We assume a MIMO-OFDM setup where users actively communicate with the anchor over multiple subcarriers with multiple antenna. The anchor location is given by $\vp_0$. Each user estimates the channel and gets \ac{CSI} values, represented as complex $N_{sc}\times N_{tx}\times N_{rx}$ arrays with $N_{sc}$ number of subcarriers, and $N_{tx}$ and $N_{rx}$ respectively number of transmit and receive antenna.  Multi-path information like \acp{ToF} and \acp{AoA} of paths are encoded in \ac{CSI} values. However, the used bandwidth and antenna length put a limit on the resolvability of paths. Nonetheless, each user can extract a set of  \acp{ToF} and \acp{AoA} from \ac{CSI} using algorithms like MUSIC and ESPIRIT  {\cite{32276}}. Denote the number of resolvable path for the user $u$ as $N_{u}$. In this paper, we focus only on the SISO case $N_{tx}=N_{rx}=1$. This means that users can only get \ac{ToF} information from \ac{CSI} samples. To be more precise, without any synchronization between users and the anchor, we can hope to only get \ac{TDoA} of each path with respect to the \ac{LOS} path in a robust way. We consider both \ac{TDoA} and \ac{ToF} case in this paper. We assume that a dataset of \ac{CSI} values is given without any location information available. Note that \ac{CSI} is computed anyway as part the communication chain. Therefore, compared to supervised fingerprinting positioning,  getting unlabeled \ac{CSI} values has a small overhead. Denote the CSI corresponding to the user $u\in\gU$ by $CSI_u$. The problem of joint localization and mapping can be stated as follows. Given $CSI_u$, we would like to find the environment map and the location of each user $\vp_u$. We will show that it is possible to solve this problem up to isometry, if there are enough multi-path components available. As we will see next, each multi-path yields a new anchor, called a virtual anchor, which is used to parametrize the environment.

\subsection{Environment mapping}
\begin{figure}
    \centering
    \includegraphics[width=0.49\textwidth]{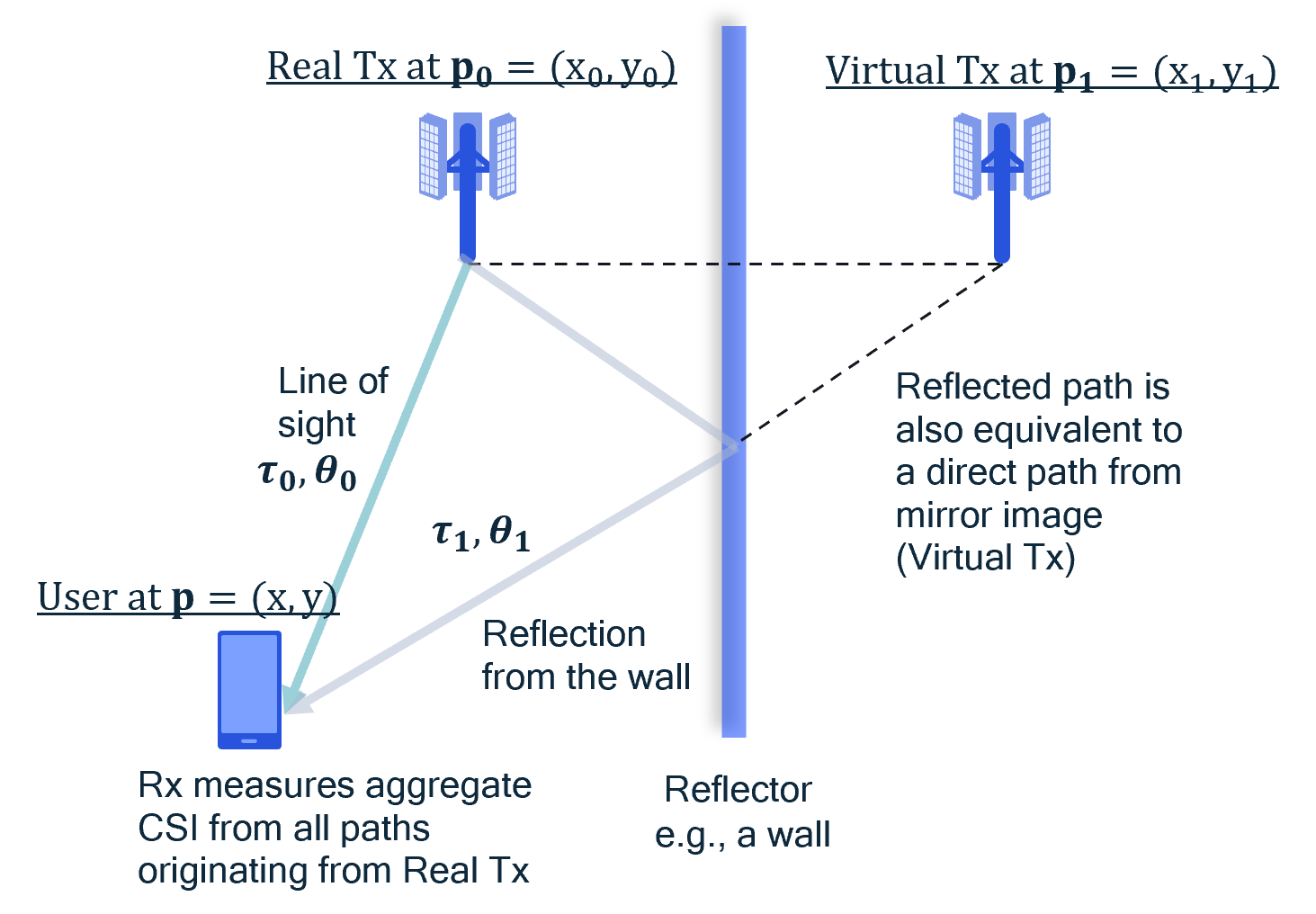}
    \caption{Virtual transmitter}
    \label{fig:va}
\vspace{-5mm}
\end{figure}

Similar to other works on SLAM \cite{6817908,gentner_multipath_2016}, the environment map from wireless propagation perspective is parametrized using virtual anchors. This means that, from user's perspective, each path originates from an anchor, which is virtually introduced as a function of the environment map and the anchor's location. 
For specular reflection, the virtual anchor position is independent of user location. However, using virtual anchors for diffraction and diffuse multi-path is more challenging. For specular reflection, the location of the virtual anchor is the reflection of anchor location from the reflector (Figure \ref{fig:va}).  For the virtual anchor location $\vp_i$ and the user location $\vp_u$,  the ToF  of the reflected path is computed as $\frac{\norm{\vp_i-\vp_u}}{c}$, similar to a physical anchor, where $c$ is the speed of light. The same holds for AoA computation. Therefore, all specular reflections in an environment can be modelled using a set of fixed virtual anchors. 

In this work, we focus only on specular reflection.  With this choice, the problem of joint localization and mapping can be probabilistically formulated as:
\begin{equation}
    \max_{\vp_i,\vp_u} \sP\left(\vp_i, i\in\gI,\vp_u, u\in \gU\vert CSI_u, u\in\gU\right),
\end{equation}
where $\gI$ is equal to $\{0,1,2,\dots,M\}$ with $M$ denoting the number of parametrized virtual anchors. In the next section, we will see how this problem can be solved using a neural network architecture. Note that in the above formulation, the anchor location $\vp_u$ is also unknown.

\section{ Neural RF SLAM}
\label{sec:neural_rf_slam}

As a surrogate for the above likelihood optimization, we minimize the difference between the ToFs that are extracted from $CSI_u$, namely $\{\tau_{j,u}\}$ and those that are reconstructed from $\vp_u$ and $\vp_i$, namely $\{\frac{\norm{\vp_i-\vp_u}}{c}, i\in\gI\}$.
We can use set-difference losses used in set prediction neural networks for example Hausdorff or Chamfer loss. The set prediction nature of the problem is related to the association problem in multi-path aided positioning. To explain this further, suppose that the user $u$ extracts $\tau_{u,j}$ from $CSI_u$. Even if the virtual anchor locations are known, the user cannot associate ToFs with virtual anchors with the exception of line of sight ToF. This challenge is known in multi-path aided positioning  {\cite{8250069}}. 

In this work, we cast the association problem of the CSI-extracted ToFs and reconstructed ToFs as an assignment problem   {\cite{kuhn1955hungarian}}. The assignment problem is about associating the elements of two sets such that the overall association cost is minimized. The cost of associating 
 the CSI-extracted ToF $\tau_{j,u}$ and the reconstructed ToF $\{\frac{\norm{\vp_i-\vp_u}}{c}, i\in\gI\}$ is simply their absolute difference. This is a combinatorial optimization problem, which can be solved in polynomial time using Hungarian algorithm \cite{kuhn1955hungarian}. The Hungarian algorithm associates $j$'th ToF to $\pi_{H}(j)$'th reconstructed ToF where $\pi_{H}(\cdot)$ is the association function between indices. The associated ToFs are two vectors that can be compared using any norm, say $\ell(\cdot,\cdot)$ on the respective vector space. The optimization problem can be written as follows:
\begin{equation*}
\min_{\vp_i,\vp_u} \sum_{u\in\gU} \ell\left(
\left(\tau_{j,u}\right)_{j\in[N_{u}]},  \left(\frac{\norm{\vp_{\pi_H({j})}-\vp_u}}{c}\right)_{j\in[N_{u}]}
\right).
\end{equation*}
One challenge of  this formulation is that any new sample in the dataset will add a new unknown $\vp_u$ to the problem with 2 or 3 unknown parameters. This makes both the optimization problem and inference for a new user challenging. However, we can amortize the inference cost by replacing $\vp_u$'s with a neural network $f_\theta(\cdot)$ that maps $CSI_u$ to the location information. Therefore, the final optimization problem is given by
\begin{gather}
\min_{\vp_i,\theta}
\sum_{u\in\gU} \ell\left(
\left(\tau_{j,u}\right)_{j\in[N_{u}]},  \left(\frac{\norm{\vp_{\pi_H({j})}-f_\theta(CSI_u)}}{c}\right)_{j\in[N_{u}]}
\right),
\end{gather}
where $\theta$ is the neural network parameter. We can solve this problem using gradient descent optimization. Once the model is trained, $f_\theta(\cdot)$ can be used as positioning function and thereby reduces the inference time significantly.







\subsection{Architecture: Neural RF SLAM}
\begin{figure}
    \centering
    \includegraphics[width=0.49\textwidth]{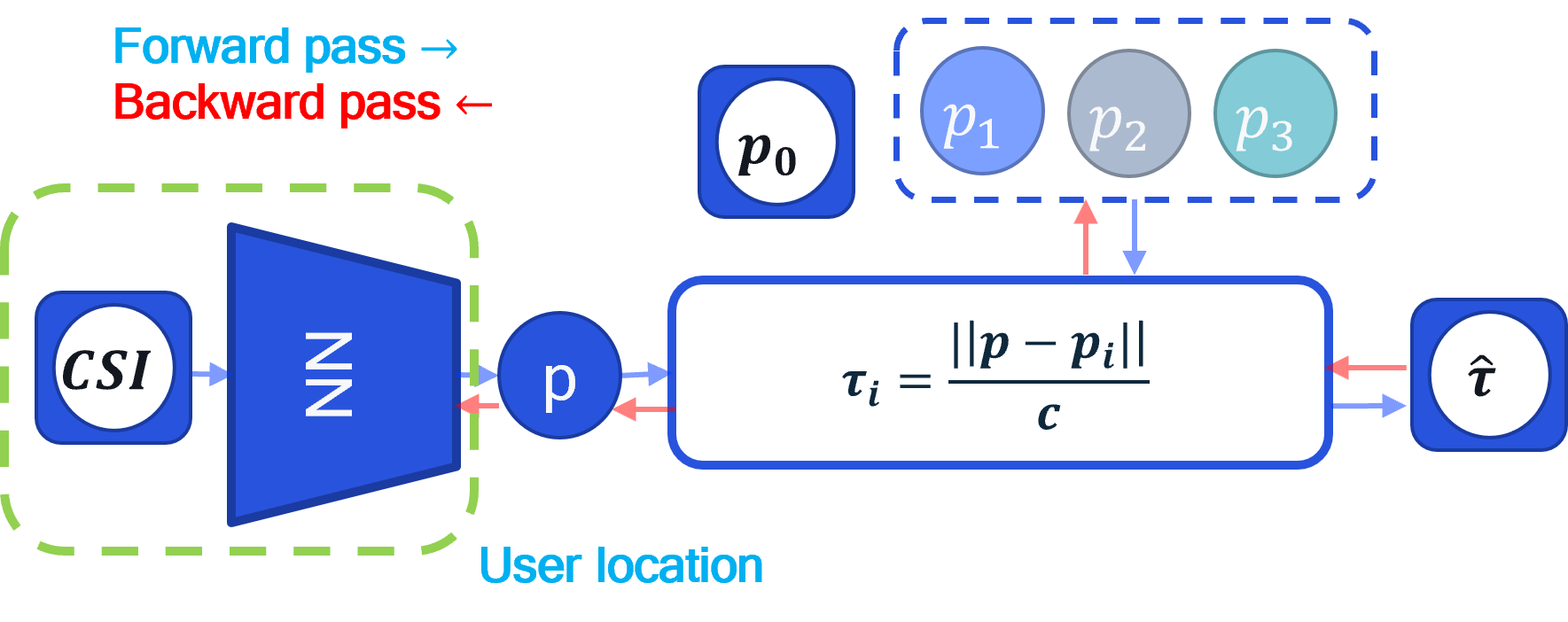}
    \caption{Neural RF SLAM}
    \label{fig:neural_slam}
    \vspace{-5mm}
\end{figure}

To summarize, we introduce an encoder-decoder architecture to solve the joint localization and mapping problem. The encoder, $f_\theta(\cdot)$ maps CSI to the user position while the decoder which is built on the physics of propagation  maps the user positions to predicted ToFs. The decoder is parametrized by Cartesian co-ordinates of virtual anchors, which are learned jointly with the encoder parameters. See Figure \ref{fig:neural_slam}. The decoder in figure \ref{fig:neural_slam} maps the output $\vp_u$ of the encoder $f_\theta(\cdot)$ to a set of ToF $\{\tau_u\}$ corresponding to multipaths from each learnt virtual anchor $\vp_j$. The decoder is randomly initialised with fixed number of learnable virtual anchors, $N_u$,  that are optimized using the objective \ac{ToF} reconstruction error. The parameter $N_u$ is selected during feature extraction (see \ref{subsec: Feature extraction}), such that the value is representative of total number of virtual anchors present in the environment. Each user lies in coverage area of a subset of all the virtual anchors $N_u$, and data association  (see \ref{subsec:data_association}) stage helps to identify the right subset. Since we optimize the ToFs with no additional information on the the size of the room or the location of anchor, the solution will have isometric ambiguities. For instance the translation and rotation of the whole space will change the coordinates but will not affect ToFs. The first impact of this ambiguity is that the optimization is harder to converge. We comment on some of the challenges in the next subsections.


 \subsection{Data Association }
 \label{subsec:data_association}
As mentioned above, for the optimization, we need to solve the set prediction task where we compute the loss between reconstructed ToF and the CSI-extracted ToF.
We use the Hungarian algorithm to associate ToFs and then calculate the loss between two vectors. We have chosen the smooth L1-loss as it gave better performance than MSE loss. We have also explored standard loss functions like Chamfer loss and greedy loss. To compute the greedy loss, we sort the ToFs and then calculate the difference of the vectors using a metric of choice. Note that we can associate line-of-sight ToFs. The smallest extracted ToFs should be associated to the main anchor with location $\vp_0$. Therefore, we can decompose the loss function into two parts, one that computes difference between line-of-sight ToFs, and the other part that needs to solve the association first and compute the difference. We have observed that this decomposition {improves} the convergence of the algorithm.

       

\subsection{Non-CSI based inputs with variable size}
The choice of CSI as input to the neural network is arbitrary. What is crucial for the model is having access to ToFs, or other modalities, that can be computed by the decoder. The input can be any modality that is correlated with the location. An obvious choice is to use the extracted ToFs. In this case, the encoder network should be a set-function that acts on the set of ToFs and is invariant to  permutation of the input. Besides, it should be able to deal with variable size inputs. To deal with variable sized sets as input, we implemented DeepSet \cite{zaheer_deep_2017} based architecture. DeepSet model first maps every single ToF to a fixed feature. The features are aggregated in a permutation invariant way for example by averaging, and passed to a second neural network. Note that CSI can be seen as a permutation invariant embedding of ToFs.

\subsection{Feature extraction and virtual anchor initialization}
\label{subsec: Feature extraction}
When the genie ToF information is not available, we rely on classical super-resolution based algorithms to get them from CSI. MUSIC \cite{32276} algorithm uses the frequency response of \ac{CIR} to compute the eigen- directions of source and noise components. Further for better performance in noisy environment, MUSIC algorithm uses multiple CSI measurements per user location to compute a stochastic covariance matrix. 
To solve source enumeration and thus approximate the number of virtual anchors we use Minimum Description length (MDL) approach given in \cite{barron1998minimum}. 
We can use a larger number of virtual anchors during the training time. As the training progresses, only a subset of them continue to update, and the rest will not be updated and can be removed post-training. We have observed that this over-parametrization helps the optimization.



\subsection{Resolving isometric ambiguities}
We have mentioned that the problem can be solved only up to isometry (See Figure \ref{fig:slam_example}). In other words, the solution for ToF-based decoder and unknown anchor and virtual anchor locations remains valid after rotation, translation and reflection. The presence of these isometries in the solution space make the optimization problem challenging. We can break some of these symmetries. For example, we can always arbitrarily fix the anchor location, which breaks the translation symmetry. 
Further for one of the virtual anchors, we can fix all its coordinates except one to the same coordinates chosen for the anchor location. For example, in 2D case, we fix the anchor location to $(0,0)$ and one virtual anchor to $(0,y)$. In this way, we can partially break the rotation symmetry, because only those rotations yield a valid solution that map the coordinates of one virtual anchor to the fixed coordinate.
The remaining ambiguities are reflections and these rotations. These ambiguities can be fixed post-training using few samples with known location to learn the isometries. These samples are only needed to learn the rotation matrix. Note that the core network is still trained in an unsupervised way, and the labeled data is only used for the correction network.

\subsection{Other modalities and multi-anchor case}
We have focused so far on ToF as the main modality. Although TDoA, between LoS path and reflected paths, would have at least the same ambiguities as ToF, adding AoA can break reflection symmetries. Throughout this paper, our focus is on ToF and TDoA. We assume SISO communication, which means that we cannot extract AoA from CSI.
Finally, the model can be extended to multi-anchor by concatenating multiple decoders, one per anchor. 

\subsection{Non-specular Multipaths}
Although the we focused so far on dealing with multiple bounces in the environment, Neural SLAM decoder is well poised to reconstruct other propagation effects that can be modeled using geometrically consistent ray tracing approach. When the environment consists of diffuse multi-paths some of received ToFs do not provide additional information on the locations of virtual transmitters and these paths generally appear to form unresolvable paths occurring around one delay element of impulse response. Some works model this using hybrid geometric models that constitute two kinds of components where non-specular multipaths are considered interfering components with no underlying geometrical correlation with the transmitter location. In recent works, such as \cite{wen20205g}, the diffuse scattering is modelled using a cluster where the cluster spread is dependent on the smoothness of the wall, while some have approached this by treating NLOS paths due to diffraction as noise. 

Ray tracing approaches have tried to model the diffracted path using geometric properties and material properties of wedge. One such approach is geometric theory of diffraction (GTD) \cite{keller1962geometrical}. GTD assumes that for UE locations that lie in the shadow, the diffracted rays form a “Kellers cone” with the edge as the common axis and the incidence point as the apex of the cone \cite{yun2015ray}.  As the UE moves around the shadow region the diffracted ray appears to have originated from a point of incidence which moves along the edge. While the virtual anchor from which it appears to originate can be established by 3D rotation of true anchor \cite{mucalo2013virtual}. GTD based ray-tracing models are good approximation to the real world scenarios as reported in \cite{rappaport2017small}. It is shown that in an indoor environment, the knife edge diffraction models overestimate the losses for certain polarizations , incident angles and wall properties. This problem can be circumvented using GTD models for other scenarios with irregular shaped edges. Note that the above results are for frequencies above 6 Ghz. 

The GTD model provides a geometric framework to model time of flights due to diffraction. This can be incorporated in the decoder part of the network by using a virtual anchor location parametrization that is dependent on the user location. We leave these extensions as future works.

\begin{figure}
    \centering
    \includegraphics[width=0.49\textwidth]{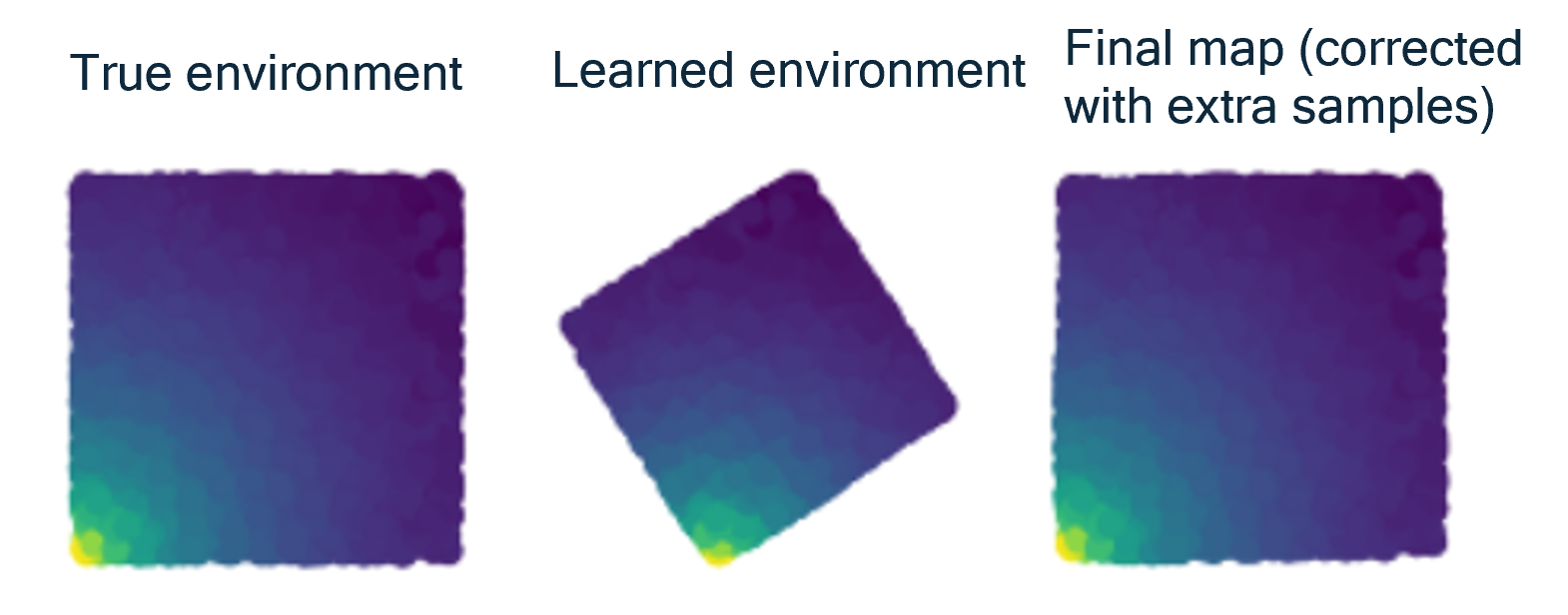}
    \caption{Learning can be done up to isometries}
    \label{fig:slam_example}
    \vspace{-5mm}
\end{figure}
\section{Experiments}
\label{sec:experiments}

We implement a simple MLP, when ToFs are used as input to the positioning network. When CSI is used as input, we use a convolutional architecture  with convolutional layer, batchnorm and RelU non-linearity.  We have also implemented DeepSet \cite{zaheer_deep_2017} based architecture to deal with the variable sized ToF inputs. We train these networks using Adam optimizer.For the main experiments, we have used \ac{TDoA} extracted from \ac{CSI} using MUSIC as the most challenging scenario. For ablation studies, to isolate the impact of feature extraction, we have sometimes used ground truth \ac{ToF} or \ac{TDoA}.
 
 \subsection{Datasets}
 
\begin{figure}%
    \centering
    \subfloat[\centering 3D Floor plan]{{\includegraphics[ width=3.6cm]{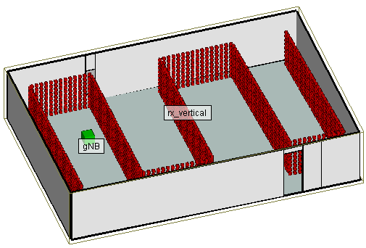} }}%
    \qquad
    \subfloat[\centering single bounce predicted point cloud]{{\includegraphics[height=3cm, width=0.5\linewidth]{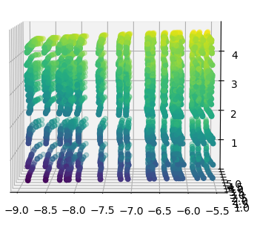} }}%
    \subfloat[\centering double bounce predicted point cloud]{{\includegraphics[height=3cm, width=0.5\linewidth]{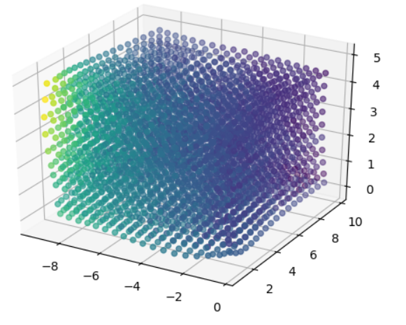} }}%
    \qquad
    \subfloat[\centering CDF of localization error 3D single bounce  ]{{\includegraphics[height=4cm ,width=0.5\linewidth]{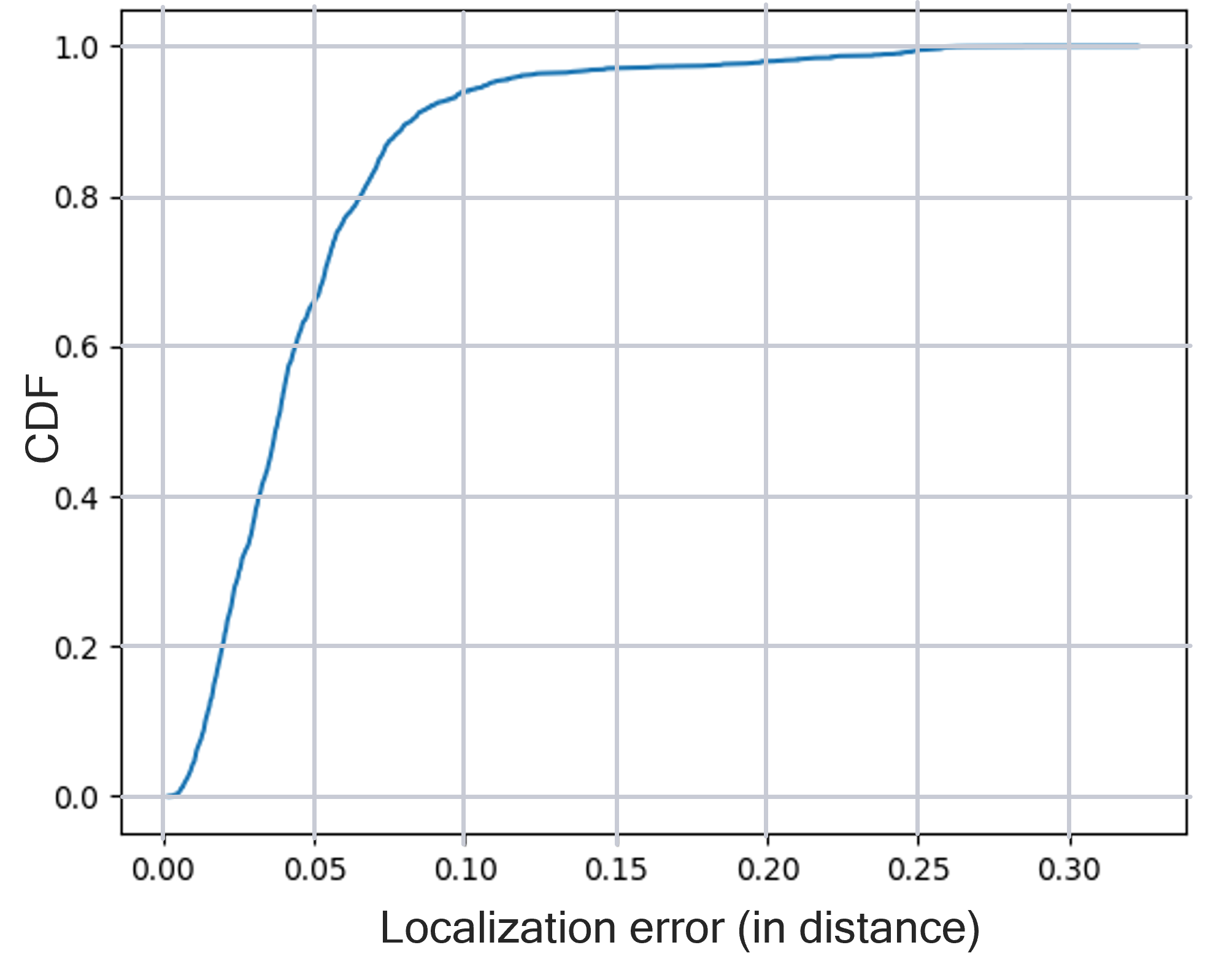} }}%
    \subfloat[\centering CDF of localization error 3D double bounce ]{{\includegraphics[height=4cm ,width=0.5\linewidth]{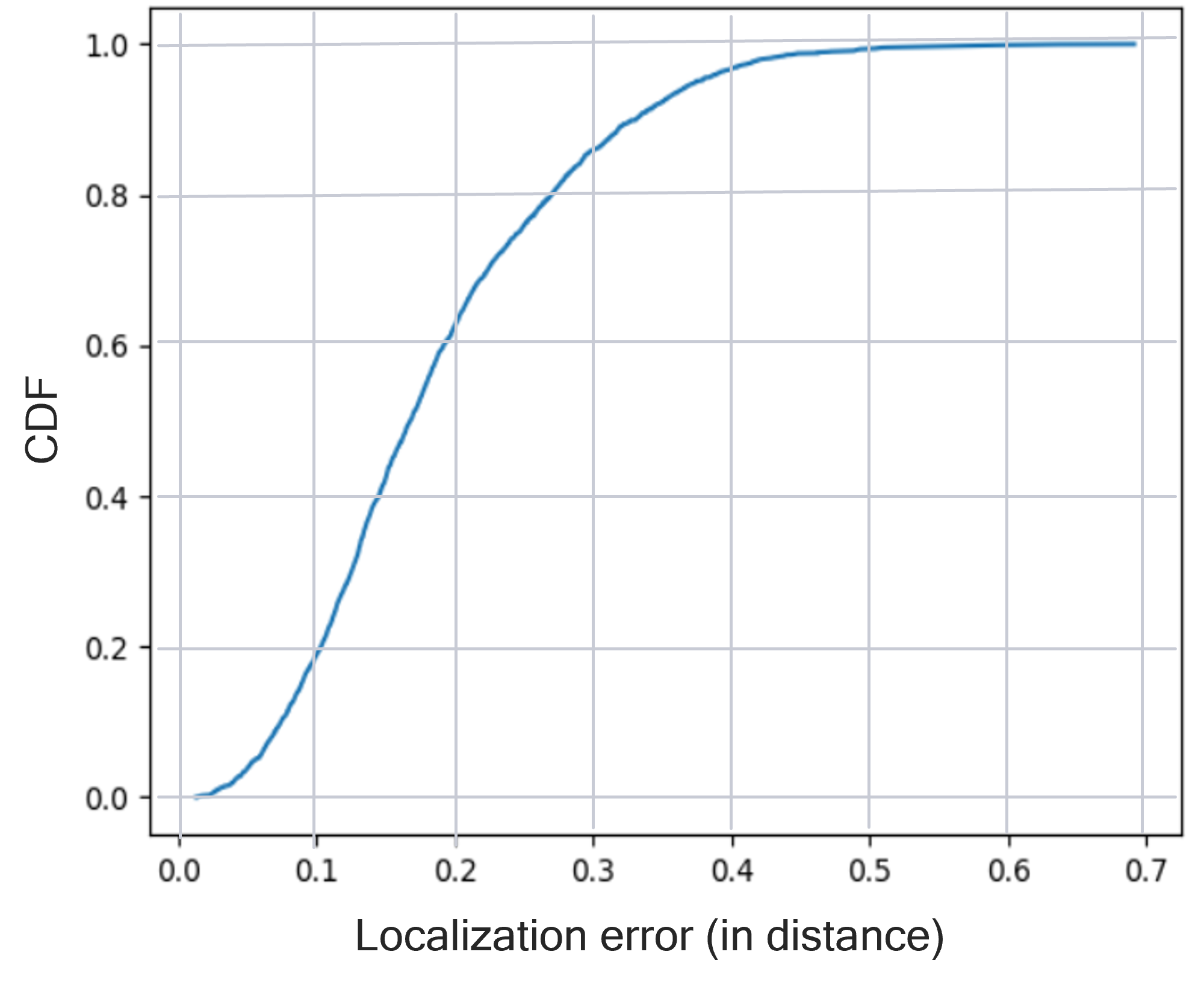} }}%
    \caption{3D TDoA Unsupervised end to end training}%
    \label{fig:3D  ToF Neural SLAM Visualizations for 6 VA and 25 VA case }%
    \vspace{-3mm}
\end{figure}

  We have used two datasets for a 2D and 3D environment. We assumed an OFDM based SISO communication with the parameters shown in table ~\ref{tab: System Params}. CSI values for this communication system are used to extract ToF and TDoA  using MUSIC algorithm. We have implemented a simple ray tracer for reflection-based environment from which the 2D dataset is generated. The environment is a room with 4 reflecting surfaces, and the transmitter placed at a corner located at $(0.1, 0.1)$.   
 For the 3D dataset \ref{fig:3D  ToF Neural SLAM Visualizations for 6 VA and 25 VA case } (a), we have used REMCOM software to generate data \cite{Wint} for an indoor environment. We consider a single room environment enclosed within 4 walls, a ceiling and a floor. There is a single anchor fixed located at $(-10, 0, 4)$. We assume the surfaces of the walls are smooth, and there is no diffraction or diffused scattering in the environment. We have datasets with single and double reflection bounces.  In both environments, the samples are randomly taken from the environment. The dataset consists of  4000 samples for the 2D case, and 13000 for the 3D case. We evaluate our model on both of these datasets. One tenth of samples is used for the test set.
For qualitative evaluation, we visualize the samples in the test set as a point cloud where each point is color coded by logarithm of its CSI magnitude. An example is given in Figure \ref{fig:slam_example} for a model trained with ground truth \acp{ToF}. It is clear that the model learns up to isometry. We report median and 90 \% quantile of the error.
\begin{table}[t]
\vspace{3mm}
\begin{center}
\begin{tabular}{|c|c|c|}
 \hline
 Parameter& 2D & 3D\\
 \hline
 Carrier frequency & 2Ghz& 3.5Ghz \\
 Bandwidth & 400Mhz & 100Mhz \\
 Num of carriers & 128 & 128\\
 Test Area & 5mx5m & 10mx10mx4m \\
 number of walls & 4 & 6  \\
 \hline
\end{tabular}
\end{center}
\caption{System Parameters}
\label{tab: System Params}
\vspace{-5mm}
\end{table}
\subsection{Baselines}
We compare Neural RF SLAM with four baselines. The first baseline is the MAP assisted single anchor approach in \cite{9013365}, which requires both noisy ToF and AoA information and {known map}. Another baseline is supervised fingerprinting positioning where we have access to CSI samples labelled with user location. We also consider two other variations of our model. First, we consider the case where the environment map is given, and virtual anchors are known. The next case is when the user location is given but the map is unknown. In both of these cases, our architecture can still be used to solve the problem by fixing some of the respective parameters in the optimization. For the latter two cases we train using MUSIC extracted labels. The results are reported in Table \ref{tab: baselines}.
\begin{table}[ht]
\centering
\begin{adjustbox}{max width=0.48\textwidth}
\small

\begin{tabular}{|l|c|c|c|}
 \hline
         &  mean(m) & median(m) & 90\% (m)  \\
\hline
1. Supervised genie user labels ConvNet & 0.02  &  0.018 & 0.05\\
2. Noisy Map assisted \cite{9013365} & 1.18 & 0.026 & 2.17   \\
3. Localization with known VA   & 0.08 & 0.12 & 0.2 \\
4. Supervised mapping  &  0.08 & 0.03 & 0.09 \\
\hline
 \end{tabular}

  \end{adjustbox}
   \caption{ Baselines for 2D Localization (1), (2), (3) and Mapping (4). Approach (2), (3), (4) use MUSIC extracted labels}
\label{tab: baselines}
\vspace{-5mm}
 \end{table}

\subsection{ End-to-End Unsupervised Neural SLAM }
\begin{figure}
\vspace{3mm}
    \centering
     \includegraphics[width=0.4\textwidth]{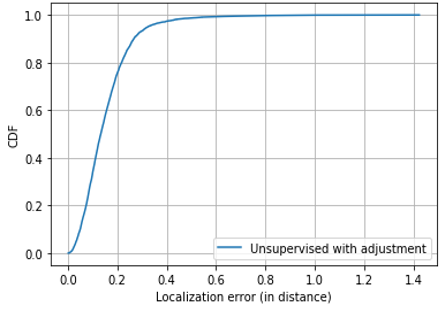}
    \caption{2D TDoA-based end-to-end unsupervised model}%
    \label{fig:TDoA-based end-to-end unsupervised model}%
\vspace{-3mm}
\end{figure}

We run Neural RF SLAM for two different modalities, namely \textit{ToF} and \textit{TDoA}. The results are as shown in tables \ref{tab: Unsupervised ToF Neural SLAM} and \ref{tab: Unsupervised TDOA Neural SLAM}. The extracted labels from CSI values for 2D case  has about 2cm of $90\%$ quantile error. In table \ref{tab: Unsupervised TDOA Neural SLAM} method (3) we extract TDoA from noisy CSI with 10dB SNR. Besides, about $10\%$ of the training data has fewer multi-paths than the number of sources in the environment. Figure \ref{fig:TDoA-based end-to-end unsupervised model} shows the CDF of localization error.  For all the experiments we report Median and $90\%$ quantile localization errors as well as mapping error reported as Median and $90\%$ quantile of  MSE between the true  and predicted virtual anchors. 
Results in table \ref{tab: Unsupervised TDOA Neural SLAM} highlight the challenge in optimizing  for TDoA for three different encoders. Note that when handling fixed size sets, MLP works well. For CSI based encoder, we use a 2 layer skip connection 1D Convolutional network and initialize the learnable virtual anchor coordinates in the decoder using samples from a random Gaussian. 


\begin{table}[ht]
\centering
\begin{adjustbox}{max width=0.5\textwidth}
\small
\begin{tabular}{|l|c|c|c|c|}
 \hline
         & Median & 90\% & Median VA & 90\% VA \\
\hline
Genie TDoA  Neural SLAM 2D & 0.03 & 0.06 & 0.01 & 0.03\\
MUSIC TDoA  Neural SLAM 2D & 0.133 & 0.26 & 0.04 & 0.08 \\
MUSIC TDoA  Noisy Neural SLAM 2D (10dB) & 0.32 & 0.45 & 0.12 & 0.18 \\
Genie TDoA  Neural SLAM 3D & 0.15 & 0.43 & 0.05 & 0.28\\
\hline
 \end{tabular}
  \end{adjustbox}
 \caption{ Unsupervised TDOA Neural SLAM}
\label{tab: Unsupervised TDOA Neural SLAM}
    \vspace{-5mm}
 \end{table}

 \begin{table}[ht]
 \vspace{3mm}
\centering
\begin{adjustbox}{max width=0.5\textwidth}
\small
\begin{tabular}{|l|c|c|c|c|}
 \hline
 \multicolumn{5}{|c|}{Single bounce Unsupervised Neural SLAM on Genie ToF } \\
 \hline
         & Median(m) & 90\%(m) & Median VA(m) & 90\% tail VA(m) \\
\hline
ConvNet  Neural SLAM 2D & 0.01 & 0.023 & 0.05 & 0.08\\
MLP  Neural SLAM 2D & 0.008 & 0.012 & 0.01 & 0.02\\
DeepSet  Neural SLAM 2D & 0.07 & 0.12  & 0.02 & 0.09\\
ConvNet  Neural SLAM 3D & 0.034 & 0.07 & 0.02 & 0.15\\
\hline
 \end{tabular}
  \end{adjustbox}
  \caption{Unsupervised ToF Neural SLAM}
  \label{tab: Unsupervised ToF Neural SLAM}
     \vspace{-5mm}
 \end{table}

\subsection{Ablation Studies}
\subsubsection{Set loss comparison} 
\begin{figure}
    \centering
    \includegraphics[width=0.35\textwidth]{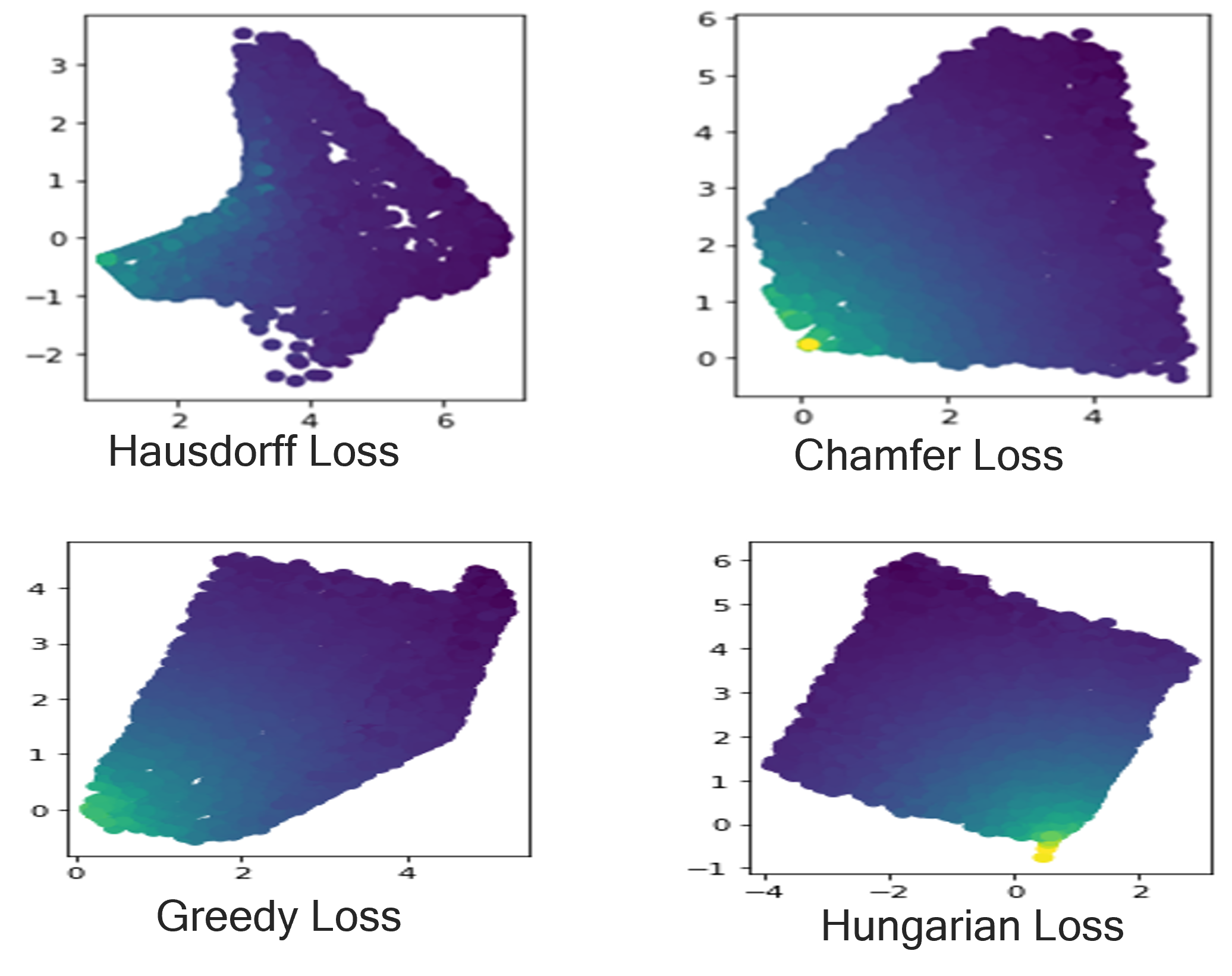}
    \caption{Set loss comparison}
    \label{fig:Set loss comparison in a 2D environment}
    \vspace{-5mm}
\end{figure}

As mentioned in earlier sections, we can use different set losses for training the network. In this part, we compare the reconstruction quality across Chamfer, Hausdorff and Hungarian loss and the greedy loss explained in Section \ref{subsec:data_association}. We trained the networks with each one of these losses once and reported the reconstructed point cloud. Figure \ref{fig:Set loss comparison in a 2D environment} shows that the Hungarian algorithm can almost perfectly recover the point cloud compared to the other losses. Note that by repeating the training multiple times and picking the one with smallest loss, we could find comparable result for all of the losses except Hausdorff loss. However, Hungarian algorithm provides better and faster convergence and does not require the overhead of multiple training rounds.  


\subsubsection{Increasing number of VA} When the dataset supports double bounce or when the number of reflecting surfaces in the  environment increase, the optimization problem involves larger variable number set prediction. We conducted experiments in 3D by introducing double bounce and increasing the number of virtual anchors from 6 to 25. For each user, some of the multi-path components have low gains and cannot be effectively extracted. This makes training more challenging, as we get fewer ToFs per user to update the virtual anchors. Table \ref{tab: Increasing the number of virtual anchors} shows the comparison when we use the ground-truth ToFs directly. Although our method still finds a solution with sub-meter accuracy, it suffers a slight performance drop when the number of virtual anchors increases.  

\begin{table}[ht]
\centering
\begin{adjustbox}{max width=0.5\textwidth}
\small
\begin{tabular}{|c|c|c|}
\hline
3D genie ToF Unsupervised SLAM & Median(m) & 90\% quantile(m) \\
\hline
6 virtual anchors & 0.034 & 0.07 \\
15 virtual anchors & 0.15  & 0.28 \\
25 virtual anchors & 0.28  & 0.40 \\
\hline
\end{tabular}
  \end{adjustbox}
  \caption{Increasing the number of virtual anchors}
   \label{tab: Increasing the number of virtual anchors}
   \vspace{-5mm}
 \end{table}
\subsubsection{Decreasing bandwidth} We also studied the effect of decreasing the bandwidth on the accuracy of 2D neural slam. Feature extraction algorithms like MUSIC resolve fewer multi-paths as the bandwidth reduces. This affects the performance of our localization network when trained on extracted ToFs resulting in optimization being stuck in a local minima. 
\begin{table}[ht]
\centering
\begin{adjustbox}{max width=0.5\textwidth}
\small
\begin{tabular}{|c|c|c|}
\hline
2D genie ToF Unsupervised SLAM & Median (m) & 90\% quantile(m) \\
\hline
100Mhz & 0.86 &  1.3\\
300Mhz & 0.15 & 0.45 \\
400Mhz & 0.13  & 0.26 \\
\hline
\end{tabular}
  \end{adjustbox}
   \label{tab: increasing the number of virtual anchors}
   \caption{Bandwidth effect on positioning}%
\vspace{-5mm}
 \end{table}
\subsection{Model complexity}
In comparison with existing deep learning based fingerprinting approaches that use resnet with 12 layers with 7x7 convolutions \cite{Ayyalasomayajula2020DeepLB}, our model uses 5x5 convolution with a single layer followed by a 2 layer MLP. Compared to nearest neighbor based models \cite{sobehy2020csi} that require larger inference time around 1.1s, the forward pass of our model takes about 3ms. Classical approaches that use majority voting based methods or least square solutions to solve the SLAM problems for every new CSI sampled from the environment require more computations. Our model does not require any optimization during inference time, once the model is trained. Table ~\ref{tab: model complexity} shows the model complexity in terms of number of trainable parameters and the forward pass model size. 
\begin{table}[ht]
\centering
\begin{adjustbox}{max width=0.5\textwidth}
\small
\begin{tabular}{|c|c|c|c|}
\hline
Model complexity (m) & number of parameters & parameter size & memory allocated\\
\hline
Convnet Neural SLAM 3D  &  156k & 580KB & 10KB \\
DeepSet Neural SLAM 3D  &  10k & 30KB &  1.7KB\\
\hline
\end{tabular}
  \end{adjustbox}
   \caption{Model complexity}
   \label{tab: model complexity}
\vspace{-5mm}
 \end{table}
\section{Conclusion}
We have presented Neural RF SLAM that can learn location and mapping up to isometry from unlabeled CSI samples. Our experiments showed that for single antenna- single anchor setup, we can achieve sub-meter accuracy in localization and mapping for specular reflection dominant environment. As future work, other propagation effects like scattering and diffraction can be added to the model. 
For some of non-specular reflection effects, it is still possible to use virtual anchor representation with ToF computed differently (see \cite{gentner_multipath_2016}). This approach seems to be more challenging for effects like diffraction. 



\bibliographystyle{IEEEtran}
\bibliography{references.bib}

\begin{thebibliography}{10}
\providecommand{\url}[1]{#1}
\csname url@samestyle\endcsname
\providecommand{\newblock}{\relax}
\providecommand{\bibinfo}[2]{#2}
\providecommand{\BIBentrySTDinterwordspacing}{\spaceskip=0pt\relax}
\providecommand{\BIBentryALTinterwordstretchfactor}{4}
\providecommand{\BIBentryALTinterwordspacing}{\spaceskip=\fontdimen2\font plus
\BIBentryALTinterwordstretchfactor\fontdimen3\font minus
  \fontdimen4\font\relax}
\providecommand{\BIBforeignlanguage}[2]{{%
\expandafter\ifx\csname l@#1\endcsname\relax
\typeout{** WARNING: IEEEtran.bst: No hyphenation pattern has been}%
\typeout{** loaded for the language `#1'. Using the pattern for}%
\typeout{** the default language instead.}%
\else
\language=\csname l@#1\endcsname
\fi
#2}}
\providecommand{\BIBdecl}{\relax}
\BIBdecl

\bibitem{art_white_0}
H.~Durrant-Whyte and T.~Bailey, ``Simultaneous localization and mapping: part
  i,'' \emph{IEEE Robotics Automation Magazine}, vol.~13, no.~2, pp. 99--110,
  2006.

\bibitem{gentner_multipath_2016}
C.~Gentner, T.~Jost, W.~Wang, S.~Zhang, A.~Dammann, and U.-C. Fiebig,
  ``Multipath {Assisted} {Positioning} with {Simultaneous} {Localization} and
  {Mapping},'' \emph{IEEE Transactions on Wireless Communications}, vol.~15,
  no.~9, pp. 6104--6117, Sep. 2016.

\bibitem{art_leitinger_BP}
E.~Leitinger, F.~Meyer, F.~Hlawatsch, K.~Witrisal, F.~Tufvesson, and M.~Z. Win,
  ``A belief propagation algorithm for multipath-based slam,'' \emph{IEEE
  Transactions on Wireless Communications}, vol.~18, no.~12, pp. 5613--5629,
  2019.

\bibitem{art_hough}
H.~Naseri and V.~Koivunen, ``Cooperative simultaneous localization and mapping
  by exploiting multipath propagation,'' \emph{IEEE Transactions on Signal
  Processing}, vol.~65, no.~1, pp. 200--211, 2017.

\bibitem{art_A035}
E.~Leitinger, P.~Meissner, C.~Rüdisser, G.~Dumphart, and K.~Witrisal,
  ``Evaluation of position-related information in multipath components for
  indoor positioning,'' \emph{IEEE Journal on Selected Areas in
  Communications}, vol.~33, no.~11, pp. 2313--2328, 2015.

\bibitem{art_A024}
Y.~Ge, H.~Kim, F.~Wen, L.~Svensson, S.~Kim, and H.~Wymeersch, ``Exploiting
  diffuse multipath in 5g slam,'' \emph{in Proc. IEEE GLOBECOM 2020}, pp. 1--6,
  2020.

\bibitem{art_A047}
H.~Kim, K.~Granström, L.~Gao, G.~Battistelli, S.~Kim, and H.~Wymeersch, ``5g
  mmwave cooperative positioning and mapping using multi-model phd filter and
  map fusion,'' \emph{IEEE Transactions on Wireless Communications}, vol.~19,
  no.~6, pp. 3782--3795, 2020.

\bibitem{art_A079}
F.~Herranz, A.~Llamazares, E.~Molinos, and M.~Ocaña, ``A comparison of slam
  algorithms with range only sensors,'' \emph{in Proc. IEEE International
  Conference on Robotics and Automation (ICRA)}, pp. 4606--4611, 2014.

\bibitem{art_A080}
B.~Boots and G.~Gordon, ``A spectral learning approach to range-only {SLAM},''
  in \emph{International {Conference} on {Machine} {Learning}}.\hskip 1em plus
  0.5em minus 0.4em\relax PMLR, 2013, pp. 19--26.

\bibitem{art_A081}
J.~Djugash and S.~Singh, ``A robust method of localization and mapping using
  only range,'' in \emph{Experimental {Robotics}}.\hskip 1em plus 0.5em minus
  0.4em\relax Springer, 2009, pp. 341--351.

\bibitem{Ayyalasomayajula2020DeepLB}
R.~S. Ayyalasomayajula, A.~Arun, C.~Wu, S.~Sharma, A.~R. Sethi, D.~Vasisht, and
  D.~Bharadia, ``Deep learning based wireless localization for indoor
  navigation,'' \emph{Proceedings of the 26th Annual International Conference
  on Mobile Computing and Networking}, 2020.

\bibitem{32276}
R.~Roy and T.~Kailath, ``Esprit-estimation of signal parameters via rotational
  invariance techniques,'' \emph{IEEE Transactions on Acoustics, Speech, and
  Signal Processing}, vol.~37, no.~7, pp. 984--995, 1989.

\bibitem{6817908}
C.~Gentner and T.~Jost, ``Indoor positioning using time difference of arrival
  between multipath components,'' in \emph{International Conference on Indoor
  Positioning and Indoor Navigation}, 2013, pp. 1--10.

\bibitem{8250069}
M.~Ulmschneider, C.~Gentner, T.~Jost, and A.~Dammann, ``Multiple hypothesis
  data association for multipath-assisted positioning,'' in \emph{2017 14th
  Workshop on Positioning, Navigation and Communications (WPNC)}, 2017, pp.
  1--6.

\bibitem{kuhn1955hungarian}
H.~W. Kuhn, ``The hungarian method for the assignment problem,'' \emph{Naval
  research logistics quarterly}, vol.~2, no. 1-2, pp. 83--97, 1955.

\bibitem{zaheer_deep_2017}
M.~Zaheer, S.~Kottur, S.~Ravanbakhsh, B.~Poczos, R.~R. Salakhutdinov, and A.~J.
  Smola, ``Deep {Sets},'' \emph{Advances in Neural Information Processing
  Systems}, vol.~30, 2017.

\bibitem{barron1998minimum}
A.~Barron, J.~Rissanen, and B.~Yu, ``The minimum description length principle
  in coding and modeling,'' \emph{IEEE transactions on information theory},
  vol.~44, no.~6, pp. 2743--2760, 1998.

\bibitem{wen20205g}
F.~Wen, J.~Kulmer, K.~Witrisal, and H.~Wymeersch, ``{5G} positioning and
  mapping with diffuse multipath,'' \emph{IEEE Transactions on Wireless
  Communications}, vol.~20, no.~2, pp. 1164--1174, 2020, publisher: IEEE.

\bibitem{keller1962geometrical}
J.~B. Keller, ``Geometrical theory of diffraction,'' \emph{Josa}, vol.~52,
  no.~2, pp. 116--130, 1962.

\bibitem{yun2015ray}
Z.~Yun and M.~F. Iskander, ``Ray tracing for radio propagation modeling:
  Principles and applications,'' \emph{IEEE Access}, vol.~3, pp. 1089--1100,
  2015.

\bibitem{mucalo2013virtual}
A.~K. Mucalo, R.~Zentner, and T.~Dela{\v{c}}, ``Virtual source modeling for
  diffraction in reference channel models,'' in \emph{2013 7th European
  Conference on Antennas and Propagation (EuCAP)}.\hskip 1em plus 0.5em minus
  0.4em\relax IEEE, 2013, pp. 1880--1883.

\bibitem{rappaport2017small}
T.~S. Rappaport, G.~R. MacCartney, S.~Sun, H.~Yan, and S.~Deng, ``Small-scale,
  local area, and transitional millimeter wave propagation for 5g
  communications,'' \emph{IEEE Transactions on Antennas and Propagation},
  vol.~65, no.~12, pp. 6474--6490, 2017.

\bibitem{Wint}
Wireless-insite.
  \url{https://www.remcom.com/wireless-insite-em-propagation-software/}.

\bibitem{9013365}
O.~Kanhere, S.~Ju, Y.~Xing, and T.~S. Rappaport, ``Map-assisted millimeter wave
  localization for accurate position location,'' in \emph{2019 IEEE Global
  Communications Conference (GLOBECOM)}, 2019, pp. 1--6.

\bibitem{sobehy2020csi}
A.~Sobehy, E.~Renault, and P.~M{\"u}hlethaler, ``Csi-mimo: K-nearest neighbor
  applied to indoor localization,'' in \emph{ICC 2020-2020 IEEE International
  Conference on Communications (ICC)}.\hskip 1em plus 0.5em minus 0.4em\relax
  IEEE, 2020, pp. 1--6.

\end{thebibliography}
\end{document}